\documentclass[a4paper,10pt,twoside]{cpc-hepnp}

\usepackage{multicol}
\usepackage{graphicx}
\usepackage{booktabs}
\usepackage{amssymb,bm,mathrsfs,bbm,amscd}
\usepackage[tbtags]{amsmath}
\usepackage{lastpage}

\begin{document}

\fancyhead[co]{\footnotesize L. S. Geng et al: The lowest-lying spin-1/2 and spin-3/2 baryon magnetic moments in
chiral perturbation theory}

\footnotetext[0]{Received 15 December 2009}

\title{The lowest-lying spin-1/2 and spin-3/2 baryon magnetic moments in
chiral perturbation theory}

\author{%
      L.S. Geng$^{1;1)}$\email{lsgeng@ific.uv.es}%
\quad J. Martin-Camalich$^{1;2)}$\email{camalich@ific.uv.es}%
\quad L. Alvarez-Ruso$^{2;3)}$\email{alvarez@teor.fis.uc.pt} \quad
M. J. Vicente-Vacas$^{1;4)}$\email{vicente@ific.uv.es} } \maketitle

\address{%
1~( Departamento de F\'{\i}sica Te\'orica and IFIC, Universidad de
Valencia-CSIC, E-46071 Valencia, Spain)\\
2~(Centro de F\'isica Computacional, Departamento de F\'isica, Universidade de Coimbra, P3004-516 Coimbra, Portugal)\\
}

\begin{abstract}
We review some recent progress in our understanding of the
lowest-lying spin-1/2 and spin-3/2 baryon magnetic moments (MMs) in
terms of Chiral Perturbation Theory (ChPT). In particular, we show
that at next-to-leading-order ChPT can describe the MMs of the octet
baryons quite well. We also make predictions for the decuplet MMs at
the same chiral order. Among them, the MMs of the $\Delta^{++}$ and
$\Delta^+$ are found to agree well with data within the experimental
uncertainties.
\end{abstract}



\begin{multicols}{2}

\section{Introduction}
The magnetic moments (MMs) of the octet baryons have long been
related to those of the proton and neutron, i.e., the celebrated
Coleman-Glashow (CG) relations~\cite{Coleman:1961jn}. These
relations are a result of (approximate) global SU(3) flavor
symmetry. Of course, we know that SU(3) flavor symmetry is broken,
as can also be clearly seen by comparing the predicted MMs with the
corresponding experimental values (see Table 1). How to implement
SU(3) breaking in a model-independent and systematic way has been
pursued since then.

Chiral symmetry and its breaking pattern, in combination with the
concept of effective field theory first systematically put forward
by Weinberg~\cite{Weinberg:1978kz}, has led to a low-energy
effective theory of QCD--Chiral Perturbation Theory
(ChPT)~\cite{Gasser:1983yg,Gasser:1984gg,Gasser:1987rb,Bernard:1995dp,Pich:1995bw,Scherer:2002tk,Bernard:2007zu,Scherer:2009bt}.
It has long been realized that ChPT may be employed to study SU(3)
breaking effects on the MMs of the baryon octet.  The first effort
was undertaken by Caldi and Pagels in 1974~\cite{Caldi:1974ta}, even
before ChPT as we know today was formulated. It was found that at
next-to-leading-order (NLO), SU(3) breaking effects are so large
that the description of the octet baryon MMs by the CG relations
tends to deteriorate, which was later confirmed by the calculations
performed in Heavy Baryon (HB)
ChPT~\cite{Jenkins:1992pi,Durand:1997ya,Puglia:1999th,Meissner:1997hn}
and Infrared (IR) ChPT~\cite{Kubis:2000aa}. This apparent failure
has often been used to question the validity of SU(3) ChPT in the
one-baryon sector.  In order to solve this problem, different
approaches have been suggested, including reordering the chiral
series~\cite{Mojzis:1999qw}
 or using a cutoff to reduce the loop contributions, i.e., the so-called long-range regularization~\cite{Donoghue:1998bs}.

We will show in this talk that the above-mentioned apparent failure
of baryon SU(3) ChPT is caused by the power-counting-restoration
(PCR) procedure used in removing the power-counting-breaking (PCB)
terms, which are due to the large non-zero baryon masses in the
chiral limit~\cite{Gasser:1987rb}. The HB, the IR, and the
extended-on-mass-shell (EOMS)\footnote{It is necessary to stress
that the EOMS regularization scheme is nothing but a $\overline{MS}$
procedure with an additional subtraction removing the finite
power-counting-breaking pieces, if they exist.} approaches all
remove the PCB terms, but the HB and IR approaches achieve this by
also removing nominally higher-order terms in such a way that
relativity and analyticity of the loop results are lost. These
questions have been discussed quite extensively in the literature,
e.g., see Refs.~\cite{Pascalutsa:2004ga,Scherer:2009bt,Geng:2008mf}.
Once the relativity and analyticity of the loop results are properly
conserved, e.g., in the EOMS regularization scheme, it was found
that baryon SU(3) ChPT at NLO improves the CG
relations~\cite{Geng:2008mf}, contrary to the conclusions of most
previous ChPT studies performed at this order.

\begin{center}
\includegraphics[width=8cm]{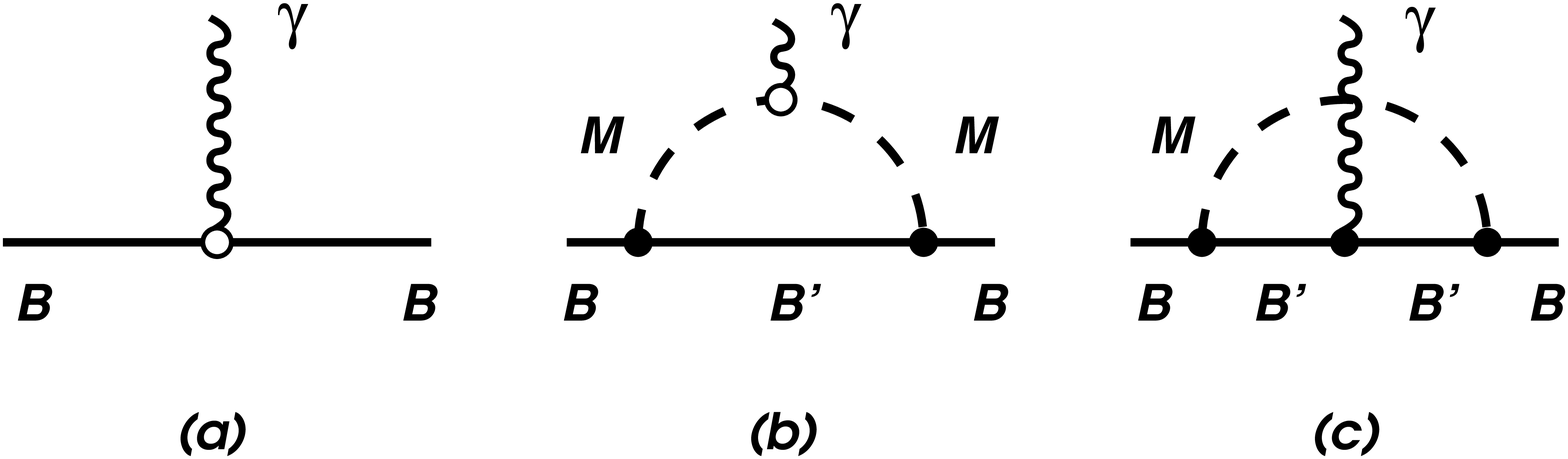}
\figcaption{\label{fig1}   Feynman diagrams contributing to the
octet baryon magnetic moments up to NLO.}
\end{center}

It has often been argued  that different regularization schemes
should yield the same results since the difference between them is
of nominal higher order. One must notice, however, that in order for
this to be true, the regularization procedure should not break the
analyticity of the loop results, which is certainly not true in
certain cases for the HB and IR schemes, as demonstrated in
Refs.~\cite{Pascalutsa:2004ga,Scherer:2009bt}. At a certain order,
one scheme can  converge faster than the other schemes. From a
practical point of view, one should better choose the one that
conserves the analyticity of the loop results, is covariant, and,
meanwhile, converges faster. Among the HB, IR, and EOMS
regularization schemes, the EOMS scheme has been found to satisfy
the above criteria. Therefore, we have chosen the EOMS
regularization scheme in all the calculations presented in this
work.

\section{The octet baryon magnetic moments}

\subsection{Dynamical octet baryon contributions}
In the following, we discuss the results for the octet baryon
magnetic moments at NLO without considering the contributions of
dynamical decuplet baryons, which will be studied in the next
sub-section.

We will not show the detailed formalism here, which can be found in Ref.~\cite{Geng:2008mf}.
Up to NLO, one has the diagrams shown in Fig.~\ref{fig1}.
 The tree-level coupling
\textbf{\textit{(a)}} gives the leading-order (LO) result
\begin{equation}
\kappa_B^{(2)}=\alpha_B b_6^D+\beta_B b_6^F, \label{eq:treeL}
\end{equation}
where the coefficients $\alpha_B$ and $\beta_B$ for each of the octet baryons
 are listed in Table I of Ref.~\cite{Geng:2008mf}. This lowest-order contribution is
nothing  but the SU(3)-symmetric prediction leading to the CG
relations ~\cite{Coleman:1961jn, Jenkins:1992pi}.

The $\mathcal{O}(p^3)$ diagrams \textbf{\textit{(b)}} and
\textbf{\textit{(c)}} account for the leading SU(3)-breaking
corrections that are induced by the corresponding degeneracy
breaking in the masses of the pseudoscalar meson octet. Their
contributions to the anomalous magnetic moment of a given member of
the octet $B$ can be written as
\begin{eqnarray}
\kappa^{(3)}_B&=&\frac{1}{8\pi^2 F_\phi^2}\left(\sum_{M=\pi,K}\xi_{BM}^{(b)}
H^{(b)}(m_M)\right.\nonumber\\
&+&\left.\sum_{M=\pi,K,\eta}\xi_{BM}^{(c)} H^{(c)}(m_M)\right)\label{eq:thirdO}
\end{eqnarray}
with the coefficients $\xi_{BM}^{(b,c)}$ listed in Table I of Ref.~\cite{Geng:2008mf}. The
loop-functions, which are convergent, read
\begin{eqnarray}
 H^{(b)}(m)&=&-M_B^2+2 m^2+\frac{m^2}{M_B^2}(2
M_B^2-m^2)\log\left(\frac{m^2}{M_B^2}\right)\nonumber \\
&+&\frac{2m\left(m^4-4 m^2 M_B^2+2
M_B^4\right)}{M_B^2\sqrt{4M_B^2-m^2}}\,\arccos\left(\frac{m}{2\,M_B}\right),
\nonumber \\
H^{(c)}(m)&=&M_B^2+2m^2+\frac{m^2}{M_B^2}(M_B^2-m^2)\log\left(\frac{m^2}{M_B^2}
\right)\nonumber \\
&+&\frac{2m^3\left(m^2-3
M_B^2\right)}{M_B^2\sqrt{4M_B^2-m^2}}\,\arccos\left(\frac{m}{2\,M_B}\right).
\label{eq:loop}
\end{eqnarray}
 One immediately notices that they contain pieces $\sim
M_B^2$ that contribute at $\mathcal{O}(p^2)$ to the MMs, which break
the naive PC.

Different  regularization schemes differ in how they remove the PCB
terms: HB performs a dual expansion while IR subtracts from the full
result the regular part. The underlying reason that one can perform
a regularization on the results shown in Eq.~(\ref{eq:loop}) lies in
the fact that ChPT includes all symmetry allowed terms such that the
PCB terms can be absorbed by the corresponding low-energy-constants
(LECs). One can easily see that the PCB terms ($\sim M_B^2$) can be
absorbed by redefining $b^D_6$ and $b^F_6$. This is how one performs
the regularization in the EOMS scheme. In the HB and IR schemes, one
also removes higher order analytic terms while those LECs
corresponding to these nominally higher-order terms are not
explicitly taken into account in the NLO calculation. One should
also notice that in order for the EOMS argument to be totally true,
one has to use a common decay constant $F_\phi$ for pions, kaons,
and etas, because one has only two LECs at his disposal at this
order, which can not take care of higher-order effects leading to
different values for $F_\phi$.

In Table \ref{table2}, we show the LO and NLO results obtained in
the EOMS scheme~\cite{Geng:2008mf}. For the sake of comparison, we
also show the NLO results obtained by using the HB and IR schemes.
To compare with the results of earlier studies, we define
\begin{equation}
 \tilde{\chi}^2=\sum (\mu_\mathrm{th}-\mu_\mathrm{exp})^2,
\end{equation}
 while $\mu_\mathrm{th}$ and $\mu_\mathrm{exp}$ are theoretical
and experimental MMs of the octet baryons. The results shown in
Table \ref{table2} are obtained by minimizing $\tilde{\chi}^2$ with
respect to the two LECs $\tilde{b}^D_6$ and $\tilde{b}^F_6$,
renormalized $b^D_6$ and $b^F_6$. It is clear that the HB and IR
results spoil the CG relations, as found in previous studies, while
the EOMS results improve them.

\end{multicols}

\begin{center}
\scriptsize
\tabcaption{The baryon-octet magnetic moments (in nuclear magnetons)
up to $\mathcal{O}(p^3)$ obtained in different $\chi$PT approaches
in comparison with data. \label{table2}}
\begin{tabular*}{160mm}{ccccccccccc}
\toprule
 & $p$ & $n$ & $\Lambda$ & $\Sigma^-$ & $\Sigma^+$ & $\Sigma^0$ & $\Xi^-$ &
$\Xi^0$ & $\Lambda\Sigma^0$ &
$\tilde{\chi}^2$ \\
\hline
\multicolumn{11}{c}{$\mathcal{O}(p^2)$}  \\
\hline
Tree level & 2.56 & -1.60 & -0.80 & -0.97 & 2.56 & 0.80 & -1.60 & -0.97 & 1.38 & 0.46  \\
\hline
\multicolumn{11}{c}{$\mathcal{O}(p^3)$}  \\
\hline
HB  & 3.01 & -2.62 & -0.42 & -1.35 & 2.18 & 0.42 & -0.70 & -0.52 & 1.68 & 1.01  \\

IR & 2.08 & -2.74 & -0.64 & -1.13 & 2.41 & 0.64 & -1.17 & -1.45 & 1.89 & 1.86 \\

EOMS & 2.58 & -2.10 & -0.66 & -1.10 & 2.43 & 0.66 & -0.95 & -1.27 & 1.58 & 0.18  \\
\hline Exp. &  2.793(0) & -1.913(0) & -0.613(4) & -1.160(25) &
2.458(10) & --- &
-0.651(3) &-1.250(14) & $\pm$  1.61(8) &   \\
\bottomrule
\end{tabular*}
\end{center}

\begin{multicols}{2}

\begin{center}
\includegraphics[width=7cm]{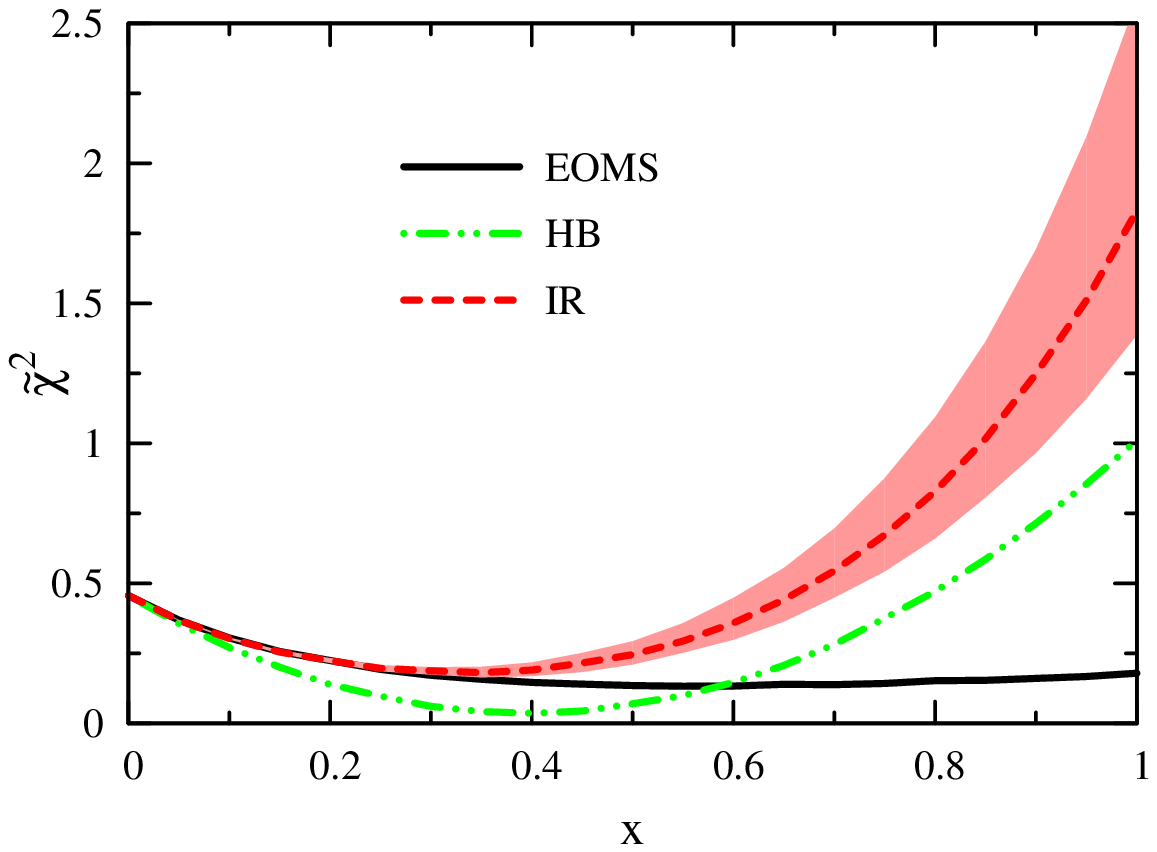}
\figcaption{\label{fig2}   SU(3)-breaking evolution of the minimal
$\tilde{\chi}^2$ in the $\mathcal{O}(p^3)$ $\chi$PT approaches under
study. The shaded bands are produced by
 varying $M_B$ from 0.8 to 1.1 GeV.}
\end{center}

The difference between the EOMS, HB, and IR approaches can also be
seen from Fig.~\ref{fig2}, where we show the evolution of the
minimal $\tilde{\chi}^2$ as a function of $x=M_M/M_{M,phys}$, while
$M_M$, $M_{M,phys}$ are the masses of the pion, kaon, eta used in
the calculation and their physical values. It is clear that at
$x=0$, the chiral limit, all the results are identical to the CG
relations. As $x$ approaches 1, where the meson masses equal to the
physical values, only the EOMS results show a proper behavior, while
both the HB and IR results rise sharply. This shows clearly that
relativity and analyticity of the loop results play an important
role in the present case.

\subsection{Dynamical decuplet baryon contributions}
Chiral perturbation theory relies on the assumption that there exists a natural cutoff such that
high-energy degrees of freedom can be integrated out, with their effects approximated by the LECs. In the case of baryon SU(3)
ChPT, the average mass gap between the baryon octet and the baryon decuplet is only 0.231 GeV, similar to the pion mass
and even smaller than the kaon mass. Therefore, in baryon SU(3) ChPT, one has to be careful about the contributions of
the decuplet baryons.

 It must be pointed out that description of spin-3/2 baryons in a
fully consistent quantum field theory framework is an unsolved problem,
see e.g. Refs.~\cite{Pascalutsa:2006up,Geng:2009hh} and references therein.

Using the ``consistent coupling'' scheme to describe the
self-interaction of spin-3/2 baryons and their interaction with
spin-1/2 baryons, we have shown in Ref.~\cite{Geng:2009hh} that the
inclusion of dynamical spin-3/2 baryons has only a small effect on
our description of the octet baryon MMs as described above. It is
also shown that this conclusion is stable with respect to all of the
model parameters within their uncertainties~\cite{Geng:2009hh}.

\section{The decuplet baryon magnetic moments}
In recent years, there have been increasing interest from both the experimental side and the lattice
QCD community to study the magnetic moments of the lowest-lying decuplet baryons, particularly those of
the $\Delta(1232)$'s. Encouraged by the success of the baryon ChPT
in describing the octet baryon magnetic moments, we have extended the same framework to
study the decuplet baryon magnetic moments. Details of this study can be found in Ref.~\cite{Geng:2009ys}.

In Table \ref{Table:ResMDM}, we show our EOMS ChPT NLO
results~\cite{Geng:2009ys} in comparison with those of a number of
theoretical models and available data. We have fitted the only LEC
at this order by reproducing the MM of the $\Omega$. It can be
clearly seen that our results for the $\Delta^{++}$ and $\Delta^+$
agree quite well with data within the experimental uncertainties.

\end{multicols}

\begin{center}
\tiny  \tabcaption{Decuplet magnetic dipole moments (in nuclear
magnetons)
 obtained in covariant ChPT up to $\mathcal{O}(p^3)$, in comparison with
 those obtained in other theoretical approaches and data.\label{Table:ResMDM}
 }
\begin{tabular*}{162mm}{ccccccccccc}
\toprule
&$\Delta^{++}$&$\Delta^+$&$\Delta^0$&$\Delta^-$&$\Sigma^{*+}$&$\Sigma^{*0}$&$\Sigma^{*-}$&$\Xi^{*0}$&$\Xi^{*-}$&$\Omega^-$\\
\hline
\multicolumn{1}{c}{SU(3)-symm.}&4.04&2.02&0&-2.02&2.02&0&-2.02&0&-2.02&-2.02\\
\multicolumn{1}{c}{NQM~\cite{Hikasa:1992je}}&5.56&2.73&-0.09&-2.92&3.09&0.27&-2.56&0.63&-2.2&-1.84\\
\multicolumn{1}{c}{RQM~\cite{Schlumpf:1993rm}}&4.76&2.38&0&-2.38&1.82&-0.27&-2.36&-0.60&-2.41&-2.35\\
\multicolumn{1}{c}{$\chi$QM~\cite{Wagner:2000ii}}&6.93&3.47&0&-3.47&4.12&0.53&-3.06&1.10&-2.61&-2.13\\
\multicolumn{1}{c}{$\chi$QSM~\cite{Ledwig:2008es}} &4.85&2.35&-0.14&-2.63&2.47&-0.02&-2.52&0.09&-2.40&-2.29\\
\multicolumn{1}{c}{QCD-SR~\cite{Lee:1997jk}}&4.1(1.3)&2.07(65)&0&-2.07(65)&2.13(82)&-0.32(15)&-1.66(73)&-0.69(29)&-1.51(52)&-1.49(45)\\
\multicolumn{1}{c}{lQCD~\cite{Leinweber:1992hy}}&6.09(88)&3.05(44)&0&-3.05(44)&3.16(40)&0.329(67)&-2.50(29)&0.58(10)&-2.08(24)&-1.73(22)\\
\multicolumn{1}{c}{lQCD~\cite{Lee:2005ds}}&5.24(18)&0.97(8)&-0.035(2)&-2.98(19)&1.27(6)&0.33(5)&-1.88(4)&0.16(4)&-0.62(1)&---\\
\multicolumn{1}{c}{large $N_c$~\cite{Luty:1994ub}}&5.9(4)&2.9(2)&---&-2.9(2)&3.3(2)&0.3(1)&-2.8(3)&0.65(20)&-2.30(15)&-1.94\\
\multicolumn{1}{c}{HB$\chi$PT~\cite{Butler:1993ej}}&4.0(4)&2.1(2)&-0.17(4)&-2.25(19)&2.0(2)&-0.07(2)&-2.2(2)&0.10(4)&-2.0(2)&-1.94\\
\hline
\multicolumn{1}{c}{ChPT~\cite{Geng:2009ys}}&6.04(13)&2.84(2)&-0.36(9)&-3.56(20)&3.07(12)&0&-3.07(12)&0.36(9)&-2.56(6)&-2.02\\
\multicolumn{1}{c}{Expt.~\cite{Yao:2006px}}&5.6$\pm$1.9&$2.7^{+1.0}_{-1.3}\pm1.5\pm3$&---&---&---&---&---&---&---&-2.02$\pm0.05$\\
\bottomrule
\end{tabular*}
\end{center}

\begin{multicols}{2}

\section{Summary and conclusions}
EOMS SU(3) baryon ChPT provides a successful description of the
SU(3) breaking effects on the octet baryon MMs. It has been found in
this particular case that the relativity and analyticity of the loop
results play an important role. We have also studied the dynamical
decuplet contributions and found that their inclusion only has a
small effect on the SU(3) breaking effects on the MMs of the octet
baryons.

Encouraged by the success of the EOMS approach, we have studied the
decuplet baryon MMs. Fitting our only LEC at this order to reproduce
the MM of the $\Omega$, we have been able to predict the MMs of the
other members of the baryon decuplet. In particular, those of the
$\Delta^{++}$ and $\Delta^+$ seem to agree well with data within the
experimental uncertainties.

This approach has also been employed to study the SU(3) breaking
corrections to the hyperon vector coupling
$f_1(0)$~\cite{Geng:2009ik}, which plays a decisive role in the
extraction of $V_{US}$ from hyperon semi-leptonic decay (HSD) data.
It will also be interesting to apply the same approach to study the
hyperon axial-vector couplings, which could provide us vital
information about the spin structure of the baryons.

\acknowledgments{This work was partially supported by the  MEC grant  FIS2006-03438 and the European Community-Research Infrastructure
Integrating Activity Study of Strongly Interacting Matter (Hadron-Physics2, Grant Agreement 227431) under the Seventh Framework Programme of EU. L.S.G. acknowledges support from the MICINN in the Program
``Juan de la Cierva.'' J.M.C. acknowledges the same institution for a FPU grant. }

\end{multicols}

\clearpage

\end{document}